\documentstyle[11pt,aaspp]{article}
\begin{document}

\title{Studying Starlight from Distant Galaxies with SIRTF}

\author{E.L.Wright (UCLA), P.Eisenhardt (JPL) \& G.Fazio (CfA)}

\begin{abstract}
Starlight from distant galaxies ($z > 3$) is redshifted into the near
infrared band with observed wavelengths from 2-8 $\mu$m.  Most of the light
is emitted by stars that have a peak emission at the 1.6 $\mu$m
wavelength of the minimum of the H$^-$ opacity.  We present
simulated images of galaxy and star fields at 3, 4.7 and 8 $\mu$m, using
the expected performance of the infrared camera on the Space InfraRed
Telescope Facility (SIRTF).
Standard astronomical image processing tools are used to locate sources,
distinguish stars from galaxies, and create color-magnitude and
color-color diagrams for the galaxies.
\end{abstract}

\section{Simulation}

The following steps have been taken in generating and analyzing 
the SIRTF images.

The galaxy population in these images was generated in clusters
with redshifts ranging from 0 to 10.  Clusters were assumed to form
with an exponential distribution of formation times having a mean
time of formation corresponding to $z=5$ in an $H_\circ = 50$, 
$\Omega = 1$ cosmology.  The density of galaxies has been increased 
slightly over the value given by Schechter in order to match Cowie's 
deep K-band counts.
Luminosities and colors evolve using Bruzual $\mu$-models with a $\mu$
that depends on galaxy type.

Stars were added to the images using the populations used by Elias.
Brown dwarfs are also included, but because we have assumed only old
cold brown dwarfs ($M = 0.05\;M_\odot, \; t = 10^{10}\;{\rm yrs}, \;
T_{color} = 462 \; {\rm K}, \;
L = 10^{-6} L_\odot$), none of the detected sources were brown dwarfs.  Of
the 32 stars in the field brighter than 0.1 $\mu$Jy at 4.7 $\mu$m, 6
are brown dwarfs, but the brightest is just fainter than the 4.7 $\mu$m
flux limit at 0.6 $\mu$Jy.  Its 8 $\mu$m flux is 1.5 $\mu$Jy which is
under the 8 $\mu$m limit as well.  Note that the color is redder than
any of the galaxies that we do detect.

Galaxy spectra are based on the spectral energy distributions for
ellipticals and spirals by Marcia Rieke, which were combined
with various weights to match Bruzual $\mu$-models in the optical.
This allowed us to include the broad IR features due to the H$^-$ opacity
minimum.  This peak at 1.6 $\mu$m rest wavelength can be seen in the
bluing of the 4.7:8 $\mu$m color at $z = 2$.


800$\times$800 pixels images covering 5.3333$^\prime$
were generated, and then
convolved with the SIRTF beam.  We assumed the diffraction limit
of an 85 cm telescope with a 30\% linear obscuration.  In order to
not exceed the SIRTF specification of 50\% encircled 
energy in a 2$^{\prime\prime}$
diameter circle, this diffraction-limited beam was further convolved
with a 1.657$^{\prime\prime}$ FWHM Gaussian.  

These 800$\times$800 pixel images with 0.39$^{\prime\prime}$
pixels were then made into
256$\times$256 pixel images with 1.17$^{\prime\prime}$
pixels by summing 3$\times$3 blocks of
pixels.  This array was sub-stepped across the field in a 3$\times$3
pattern giving a final picture with 768$\times$768 
pixels of size 1.17$^{\prime\prime}$
on 0.39$^{\prime\prime}$ centers.  
Noise was added to each pixel in this image,
using the following noise levels:
3 $\mu$m, 20.4 nJy, corresponding to 2500 seconds;
4.7 $\mu$m, 74 nJy, corresponding to 2500 seconds; 
6.2 $\mu$m, 155 nJy, corresponding to 10,000 seconds;
and 8 $\mu$m, 217 nJy, corresonding to 10,000 seconds.
Because of the 3$\times$3 substepping, a total of 9 frames at each 
wavelength
are needed, giving a total exposure time of 22500+22500+90000 seconds
or 1.6 days for the image shown.

The first step after getting the 3$\times$3 sub-stepped images is to
smooth with a 3$\times$3 box.  
These images were written out as FITS files and displayed by 
SAOIMAGE.
The RGB image was created using a FORTRAN program to convert
the intensities into a PPM (Portable Pixel Map) file which was
converted into a JPEG file using xv, and then printed to a 
HP Deskjet 1200C printer using Adobe Photoshop.
Prints were also obtained using a Kodak XL 7700 dye sublimation
printer and a Polaroid Palette film recorder.


\section{Analysis}

The 8 $\mu$m image was analyzed first.  A baseline was
subtracted from the frame.  The level was chosen by finding the
smallest range of values that covered 1/4 of the pixels in the
frame, and then taking the mean of this range.  This is an
approximate way to find the mode of an image.

With the baseline subtracted, a 7 pixel (2.73$^{\prime\prime}$)
diameter circle was
scanned across the image.  Places where the flux within this circle
reached a local maximum over at least a 5$\times$5 pixel square, 
and with fluxes
greater than 3 $\sigma$ above the baseline, were identified 
as 8 $\mu$m sources and listed.
This list was then truncated at 2 $\mu$Jy total flux, yielding 346
sources.

A similar list was constructed at 4.7 $\mu$m, truncated at 2/3 $\mu$Jy
total flux.  The source positions found from the 8 $\mu$m image were
replaced by the closest source position from the 4.7 $\mu$m image in
order to get better centroids from the higher SNR 4.7 $\mu$m image.

The flux within a $2.73^{\prime\prime}$ 
diameter circle was then found in the 3 images
at 3, 4.7 and 8 $\mu$m, giving the observed colors of the sources.
The ratio of the flux within a 5 pixel ($1.95^{\prime\prime}$) 
circle to the flux contained within $2.73^{\prime\prime}$ 
was taken on the 4.7 $\mu$m image in order to see if source size could
be used to separate stars from galaxies using the SIRTF beam.

Finally we have used the extra knowledge that comes from controlling
the simulations to generate three more data per source.  A noiseless
4.7 $\mu$m image was generated using the same beam convolution and
smoothing, as well as a noiseless image of redshift times 4.7 $\mu$m
flux.  From these images we generate the noise-free flux in a 7 pixel
circle, $F[2.73^{\prime\prime}]$,
the noiseless $F[1.95^{\prime\prime}]/F[2.73^{\prime\prime}]$
ratio, and the noiseless $(z\times F):F$ ratio,
which gives us the mean redshift of the light seen in 
$2.73^{\prime\prime}$ diameter circle centered on the source.

Looking at these mean redshifts for the stars in the image suggests
that there is about (0.5 $\mu$Jy)*($z$ unit) of redshifted 4.7 $\mu$m
flux in the $2.73^{\prime\prime}$ aperture.  
Thus these images are pushing the confusion limit.


\section{Conclusions}

The SIRTF image quality is marginal for discriminating between galaxies
and stars.  The low-$z$ clump of sources with high 
$F[1.95^{\prime\prime}]/F[2.73^{\prime\prime}]$ ratios are stars, but
several galaxies have higher ratios and thus appear more compact than some
of the stars.  Fortunately almost all the faint sources in a high latitude
deep survey will be galaxies.

Photometric redshifts using colors out to $\lambda = 8\;\mu$m
appear to be reliable.
There are 40 galaxies with $z > 3$ in the sample of 346 sources.
Sorting the 346 sources by 4.7:8 $\mu$m color, the 41 reddest
sources contain 39 out of these $z > 3$ galaxies.
The 4.7:8 $\mu$m color redshift diagram appears to be saturating at $z = 5$.  
The 3:4.7 $\mu$m color saturated at $z = 2$ and then became bluer for higher
redshifts, so the longer wavelength data is needed for reliable
photometric redshifts at high $z$.

\section{Captions}

\noindent
The $5^\prime \times 5^\prime$ 
SIRTF field of view, showing
the 8 $\mu$m image as red, the 4.7 $\mu$m image as green, and the
3 $\mu$m image as blue.

\noindent
The same region with detected sources color coded by redshift.
Dark blue for $0 < z < 1$, light blue for $1 < z < 2$, green
for $2 < z < 3$, orange for $3 < z < 4$, and red for $4 < z$.

\noindent
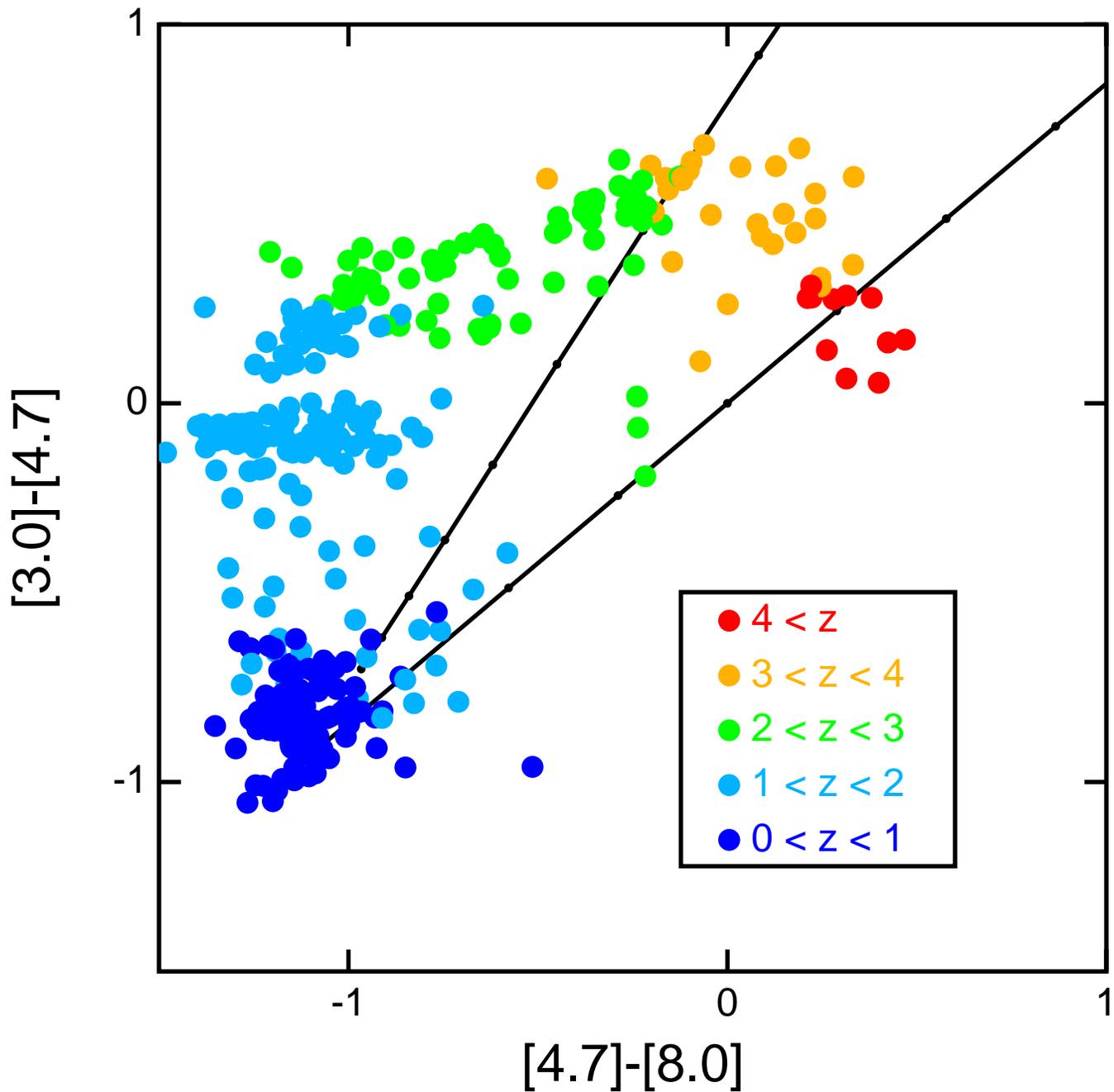
A color-color plot showing the 3:4.7 and 4.7:8 colors in magnitudes
(relative to $F_\nu$ = constant) for the detected sources, color coded
by redshift.
Dark blue for $0 < z < 1$, light blue for $1 < z < 2$, green
for $2 < z < 3$, orange for $3 < z < 4$, and red for $4 < z$.
The lines show the locations of blackbodies (upper) and power law
spectra (lower).

\noindent
A size {\it vs.} redshift plot for the detected sources.
\end{document}